\documentclass{article}
\usepackage{arxiv}
\usepackage{amsmath,amssymb}
\usepackage{graphicx}
\usepackage{dcolumn}
\usepackage{bm}

\usepackage[utf8]{inputenc}
\usepackage[T1]{fontenc}
\usepackage{mathptmx}
\usepackage{color}

\usepackage{url}
\usepackage[hidelinks,colorlinks=true, citecolor = blue]{hyperref}
\usepackage{ragged2e}
\usepackage[numbers]{natbib}
\newcommand{\apj}{{Astrophys. J.}}
\newcommand{\apjl}{{Astrophys. J. Lett.}}

\newcommand{\ssr}{{Space Sci. Rev.}}

\newcommand{\jgr}{{J. Geophys. Res.}}
\newcommand{\grl}{{Geophys. Res. Lett.}}
\newcommand{\solphys}{{Solar Physics}}

\newcommand{\aap}{{Astronomy\& Astrophysics}}

\newcommand{\mnras}{{Monthly Notices of the Royal Astronomical Society}}
\newcommand{\apjs}{{Astrophys. J. Supp.}}

\title{Particle-In-Cell Simulations of Sunward and Anti-sunward Whistler Waves in the Solar Wind}
\author{
	Ilya V. Kuzichev$^{1}$, Ivan Y. Vasko$^{2,3}$, Anton V. Artemyev$^{4,3}$, Stuart D. Bale$^{2}$, Forrest S. Mozer$^{2}$\\
	$^1$ilya.kuzichev@njit.edu\\
	$^1$New Jersey Institute of Technology, CSTR, 323 MLK Blvd, Newark, NJ 07102-1982, USA\\
	$^2$Space Sciences Laboratory, University of California, Berkeley, CA 94720-7450, USA\\
	$^3$Space Research Institute of RAS, 84/32 Profsoyuznaya Str, Moscow 117997, Russia\\
	$^4$Institute of Geophysics and Planetary Physics, University of California, Los Angeles, CA 90065-1567, USA\\
}

\begin{document}
\maketitle

\begin{abstract}
\justify
Spacecraft observations showed that electron heat conduction in the solar wind is probably regulated by whistler waves, whose origin and efficiency in electron heat flux suppression is actively investigated. In this paper, we present Particle-In-Cell simulations of a combined whistler heat flux and temperature anisotropy instability that can operate in the solar wind. The simulations are performed in a uniform plasma and initialized with core and halo electron populations typical of the solar wind. We demonstrate that the instability produces whistler waves propagating both along (anti-sunward) and opposite (sunward) to the electron heat flux. The saturated amplitudes of both sunward and anti-sunward whistler waves are strongly correlated with their {\it initial} linear growth rates, $B_{w}/B_0\sim (\gamma/\omega_{ce})^{\nu}$, where for typical electron betas we have $0.6\lesssim \nu\lesssim 0.9$. The correlations of whistler wave amplitudes and spectral widths with plasma parameters (electron beta and temperature anisotropy) revealed in the simulations are consistent with those observed in the solar wind. The efficiency of electron heat flux suppression is positively correlated with the saturated amplitude of sunward whistler waves. The electron heat flux can be suppressed by 10--60\% provided that the saturated amplitude of sunward whistler waves exceeds about 1\% of background magnetic field.  Other experimental applications of the presented results are discussed.
\end{abstract}

\section{Introduction} \label{sec:intro}
\justify

The early spacecraft measurements at 0.3--5 au \citep{Feldman75,Pilipp90,Scime94} and recent Parker Solar Probe (PSP) measurements at 0.1--0.3 au \citep{Halekas21:aa} showed that electron heat conduction in the solar wind cannot be described by the Spitzer-H\"{a}rm law \citep{Spitzer53}. The reason is that solar wind electrons are only weakly-collisional; the collisional mean free path typically exceeds the inverse gradient scale length of electron temperature in the heliosphere \citep{Salem2003,Bale13,Halekas21:aa}. In accordance with previous observations at 1 au \citep{Gary1999b,Tong18:arxiv,Tong19:apj}, PSP and Helios measurements showed that electron heat flux is bounded by a threshold dependent on local electron beta \citep{Jagarlamudi20:apj_helios,Halekas21:aa,Cattell22:apj_ww_absence}. The beta-dependent threshold indicates that wave-particle interactions are probably regulating electron heat conduction in the solar wind, and whistler waves were suggested to be the most likely wave activity involved in the regulation process \citep{Feldman76,Gary77,Scime94}. The modern spacecraft measurements have substantially advanced the understanding of electron heat conduction in the solar wind, but still have not established the heat flux regulation mechanism.

Whistler waves involved in the electron heat flux regulation can be of different origin. First, the electron heat flux can be regulated by whistler waves naturally produced by turbulence cascade \citep[][]{Vocks05,Saito&Gary07,Boldyrev19:mnras,Tang&Zank22:apj}. While early observations of broadband magnetic field fluctuations between ion and electron kinetic scales were indeed interpreted in terms of whistler waves \citep[][]{Beinroth81,Lengyel96}, the modern spacecraft measurements showed that magnetic field turbulence at these scales is dominated by kinetic Alfv\'{e}n waves \citep[][]{Chen16:jpp}. The presence of broadband whistler mode fluctuations cannot be entirely ruled out though \citep[][]{Narita16:apj} and the contribution of magnetic field turbulence to the electron heat flux regulation remains to be quantified \citep[][]{Boldyrev19:mnras}. On the other hand, whistler waves involved in the heat flux regulation can be produced by various electron-driven instabilities \citep[e.g.,][]{Gary75,Gary77,Verscharen22:frph}.

The velocity distribution function of electrons in a pristine solar wind consists of a dense thermal core population contributing about 90\% of total electron density and tenuous superthermal halo and strahl populations carrying the most of the electron heat flux \citep[e.g.,][]{Maksimovic05,Stverak09,Halekas20:apj,Salem21:arxiv}. The core and halo populations are relatively isotropic and can be described in the plasma rest frame by sunward-drifting Maxwell and anti-sunward drifting $\kappa-$distributions, respectively. In contrast, the strahl is a highly anisotropic population collimated around local magnetic field and streaming anti-sunward. It was recently suggested that oblique whistler waves driven by the strahl can potentially regulate the electron heat flux \citep{Vasko19,Verscharen19:apj}. Numerical simulations showed this instability can indeed suppress the electron heat flux by pitch-angle scattering the strahl and converting it into more or less isotropic halo \citep{Roberg-Clark19:apj,Micera20:apj,Micera21:apj}. The instability of oblique whistler waves could, in principle, explain the observed radial evolution of halo and strahl densities \citep[][]{Maksimovic05,Stverak09} as well as the observed beta-dependent threshold on the electron heat flux \citep{Halekas21:aa,Cattell22:apj_ww_absence}. However, there are currently several indications that this instability {\it does not} substantially regulate electron heat conduction in the solar wind.  

First, PSP and Helios measurements at 0.1--1 au showed the strahl parameters are statistically well below the instability threshold \citep{Jeong22a}. Second, PSP measurements at 0.1--0.5 au revealed the radial evolution of halo and strahl densities to be inconsistent with the halo being produced via pitch-angle scattering of the strahl \citep{Abraham22:apj}. Consistent with that, the recent analysis showed that halo electrons propagating sunward (almost a half of the halo population) originate in the outer heliosphere rather than evolve from the strahl \citep{Horaites22:mnras}. Third, whistler waves observed in a pristine solar wind at 0.1--1 au usually propagate within a few tens of degrees of local magnetic field \citep{Lacombe14,Stansby16,Kajdic16,Tong19:apj,Bercic21:aa,Kretz21:aa,Cattell22:apj_ww_absence}. Oblique whistler waves do present in the solar wind, but typically occur around stream interaction regions and coronal mass ejections \citep{Breneman10,Cattell21:apj}; thus, they are not likely substantially involved in the regulation of electron heat conduction in the solar wind.

The fact that whistler waves in the solar wind are typically quasi-parallel stimulates the analysis of their origin and effects. The early theoretical analysis by \cite{Gary75,Gary1994a} showed that whistler waves in the solar wind can be produced by the whistler heat flux instability (WHFI). This instability operates, when core and halo populations, isotropic or parallel-anisotropic in temperature, drift relative to each other parallel to local magnetic field; there is a heat flux parallel to the halo drift (but the net current in the plasma frame is zero) and the fastest-growing whistler waves propagate parallel to the heat flux. A strahl population that is drifting anti-sunward  does not affect the WHFI, because the unstable whistler waves are resonant only with a fraction of sunward-propagating halo electrons. The recent observations showed that the WHFI indeed operates in the solar wind and produces whistler waves propagating anti-sunward \citep{Tong19}. The recent Particle-In-Cell simulations showed that the WHFI can produce whistler waves with properties consistent with solar wind observations, but {\it cannot} regulate the electron heat flux \citep{Kuzichev19}, contrary to previous speculations \citep{Gary77,Gary1999b,Gary00}.

The fraction of whistler waves produced in the solar wind via the WHFI is still not known \citep[][]{Tong19:apj,Jagarlamudi20:apj_helios}. There are indications that whistler waves can be also produced by the instability associated with a perpendicular temperature anisotropy of the halo population \citep[][]{Vasko2020_pop,Jagarlamudi20:apj_helios}. These indications consist in statistically significant observations of the halo population with perpendicular temperature anisotropy \citep{Pierrard16,Jagarlamudi20:apj_helios,Wilson19:apj,Wilson20:apj,Salem21:arxiv} and preferential occurrence of whistler waves in association with isotropic or perpendicular anisotropic halo \citep{Tong19,Tong19:apj,Jagarlamudi20:apj_helios}. The recent reports of sunward and anti-sunward propagating whistler waves in near-Sun solar wind are also of relevance \citep{Agapitov20:apj_sunward,Mozer20:jgr}.

In this paper, we present Particle-In-Cell simulations of a combined whistler heat flux and temperature anisotropy instability that is potentially operating in the solar wind and capable of producing both sunward and anti-sunward whistler waves. We determine saturation amplitudes of the whistler waves along with their dependence on plasma parameters and demonstrate that these amplitudes can be estimated using {\it initial} linear growth rates of the whistler waves. The efficiency of this instability in electron heat flux regulation and other experimental applications of the presented results are discussed.

\section{Linear instability and simulation setup}\label{sec:setup}

We use Particle-in-Cell TRISTAN-MP code \citep{Spitkovsky08} and perform 1D3V simulations restricted to whistler waves propagating parallel and anti-parallel to background magnetic field. Ions are assumed to be an immobile neutralizing background. Electrons are represented by core and halo populations, whose initial velocity distribution functions (VDF) in the plasma frame in a non-relativistic limit, which is the case in our simulations, are described by Maxwell distributions
\begin{equation}
f_{\alpha}({\bf v})=\mathcal{N}_{\alpha}\exp\left[-\frac{m_{e}(v_{||}-u_{\alpha})^2}{2\;T_{\alpha}}-\frac{m_{e}v_{\perp}^2}{2\;A_{\alpha}\;T_{\alpha}}\right],
\label{eq:distr}
\end{equation}
where $\alpha=c,h$ correspond to core and halo populations, $v_{||}$ and $v_{\perp}$ are velocities parallel and perpendicular to background magnetic field, $\mathcal{N}_{\alpha}=n_{\alpha}A_{\alpha}^{-1}\left(m_{e}/2\pi T_{\alpha}\right)^{3/2}$ is the normalization constant, $n_{\alpha}$, $u_{\alpha}$, $T_{\alpha}$ and $A_{\alpha}$ are respectively densities, drift velocities, parallel temperatures and temperature anisotropies. The electron current is assumed to be zero, $n_{c}u_{c}+n_{h}u_{h}=0$. The electron heat flux is parallel to background magnetic field and carried predominantly by the halo population, $q_{e}\approx -n_{c}u_{c} T_{h}(3/2+A_{h})$, because the halo is several times hotter than the core population, $T_{h}/T_{c}\approx 3$--$7$ \citep[e.g.,][]{Maksimovic05,Salem21:arxiv}.


The combination of core and halo populations relatively well describes the electron VDF beyond 0.2 au, where the halo density is several times larger than the strahl density \citep{Maksimovic05,Salem21:arxiv,Abraham22:apj}. Although the halo population is better described by a $\kappa-$distribution \citep[e.g.,][]{Maksimovic05}, we consider Maxwellian halo to reduce the number of free parameters. The use of a $\kappa-$distribution would not affect the critical results of this study. The linear analysis of a combined whistler heat flux and temperature anisotropy instability shows that the growth rate normalized to electron cyclotron frequency $\omega_{ce}$ depends on the wavenumber normalized to electron inertial length $c/\omega_{pe}$ and the following parameters \citep[][]{Vasko2020_pop}
\begin{itemize}
    \item $\beta_{c}=8\pi n_{c}T_{c}/B_0^2$: core electron beta. 
    \item $n_{c}/n_0$: core density relative to total electron density, $n_0=n_{c}+n_{h}$.
    \item $u_{c}/v_{A}$: core drift velocity in units of Alfv\'{e}n speed, $v_{A}=B_0(4\pi n_0 m_p)^{-1/2}$.
    \item $T_{h}/T_{c}$: ratio of halo and core parallel temperatures.
    \item $A_{c}$ and $A_{h}$: core and halo temperature anisotropies.
\end{itemize}
Note that linear stability as well as nonlinear evolution also depend on the ratio between electron plasma and cyclotron frequencies, but this dependence is negligible once $\omega_{pe}/\omega_{ce}\gg 1$ \citep[][]{Kuzichev19,Vasko2020_pop}. In this paper, we keep $n_{c}/n_0=0.85$, $T_{h}/T_{c}=6$, $A_c = 1$, and present numerical simulations at various combinations of $\beta_{c}$, $u_{c}/v_{A}$ and $A_{h}$. The typical values of these parameters in the solar wind are $\beta_{c}=0.1$--$10$, $|u_{c}|/v_{A}=1$--$7$ and $A_{h}=1.1$--$1.5$ \citep{Pierrard16,Tong18:arxiv,Wilson19:apj,Wilson20:apj,Jagarlamudi20:apj_helios,Salem21:arxiv}. The values of these parameters used in three sets of simulations (25 runs per set) are presented in Table \ref{table}.

We performed the simulations at $\omega_{pe}/\omega_{ce}\sim 10$ that is about ten times smaller than in the realistic solar wind. More precisely, in all simulations, we assumed core electron temperature $T_{c}$ of 2 keV, and computed $\omega_{pe}/\omega_{ce}$ using the following identity, $\omega_{pe}/\omega_{ce}\equiv \left(\beta_{c} n_{0}/n_{c}\right)^{1/2}\left(m_{e}c^{2}/2T_{c}\right)^{1/2}$; for $\beta_{c}=0.3$--$3$ we have $\omega_{pe}/\omega_{ce}\approx 7$--$20$. In all simulation runs,  the length of the simulation box was $L\approx 105\;c/\omega_{ce}$ or, for $\beta_c = 1$, about $1300\;c/\omega_{pe}$. The temporal and spatial integration steps were 0.09 $\omega_{pe}^{-1}$ and $0.2\;c/\omega_{pe}$, both adequate to resolve the expected whistler waves. The number of particles per cell for each population was $4\cdot10^{4}$. We will preface the presentation of simulation results by linear stability analysis.

\begin{table}[]
    \centering
    \caption{The electron parameters for simulation sets I--III: each set consist of 25 simulation runs performed at a fixed value of core electron beta $\beta_{c}$ and 25 pairs of core drift velocity $u_{c}/v_{A}$ and halo temperature anisotropy $A_{h}$; the values of $A_{h}$ are from 1.1. to 1.5 with a step of 0.1, while the values of $u_{c}/v_{A}$ are from $-1.5$ to $-7.5$ with a step of 1.5. In all simulation runs the relative core electron density was $n_{c}/n_0=0.85$, the ratio of halo and core parallel temperatures was $T_{h}/T_{c}=6$, and the core electron population was isotropic, $A_c = 1$.}
    \begin{tabular}{|c|c|c|c|}
    \hline
    {\rm run sets} & $\beta_{c}$ & $-u_{c}/v _{A}$ & $A_{h}$ \\
    \hline
    I &  0.3  & 1.5:1.5:7.5 & 1.1:0.1:1.5,  \\
    \hline
    II &  1  & 1.5:1.5:7.5& 1.1:0.1:1.5,\\
    \hline
    III &  3  & 1.5:1.5:7.5 & 1.1:0.1:1.5,\\
    \hline
    \end{tabular}
    \label{table}
\end{table}

Figure \ref{fig1} presents results of linear stability analysis of whistler waves at fixed values of core electron beta and halo temperature anisotropy ($\beta_{c}=1$ and $A_h=1.3$), but various values of electron heat flux determined by core drift velocity $u_{c}/v_{A}$. Panels (a) and (b) present the dispersion curves and growth rates of whistler waves propagating parallel and anti-parallel to the electron heat flux. When the electron heat flux is absent ($u_{c}/v_{A}=0$), {\it identical} parallel and anti-parallel whistler waves are unstable due to the halo temperature anisotropy. The presence of electron heat flux breaks the symmetry resulting in larger growth rates of whistler waves propagating parallel to the electron heat flux. Panels (c)--(e) present the maximum growth rates along with corresponding frequencies and wave numbers of parallel and anti-parallel whistler waves unstable at various values of $u_{c}/v_{A}$. In the considered range of $u_{c}/v_{A}$ values, the parameters of the fastest-growing parallel whistler waves barely vary ($\gamma_{+}/\omega_{ce}\approx 0.01$, $\omega_{+}/\omega_{ce}\approx 0.1$ and $k_{+}c/\omega_{pe}\approx 0.34$). In contrast, the maximum growth rate of anti-parallel whistler waves monotonously decreases from $\gamma_{-}/\omega_{ce}\approx 0.01$ to $10^{-3}$; the frequency and wave number monotonously decrease by a factor of a few.

\section{Results of simulations at $\beta_{c}=1$ and $A_{h}=1.3$\label{sec:res_part}}

\begin{figure*}[ht!]
    \centering
    \includegraphics[width=1\textwidth,height=0.45\textheight]{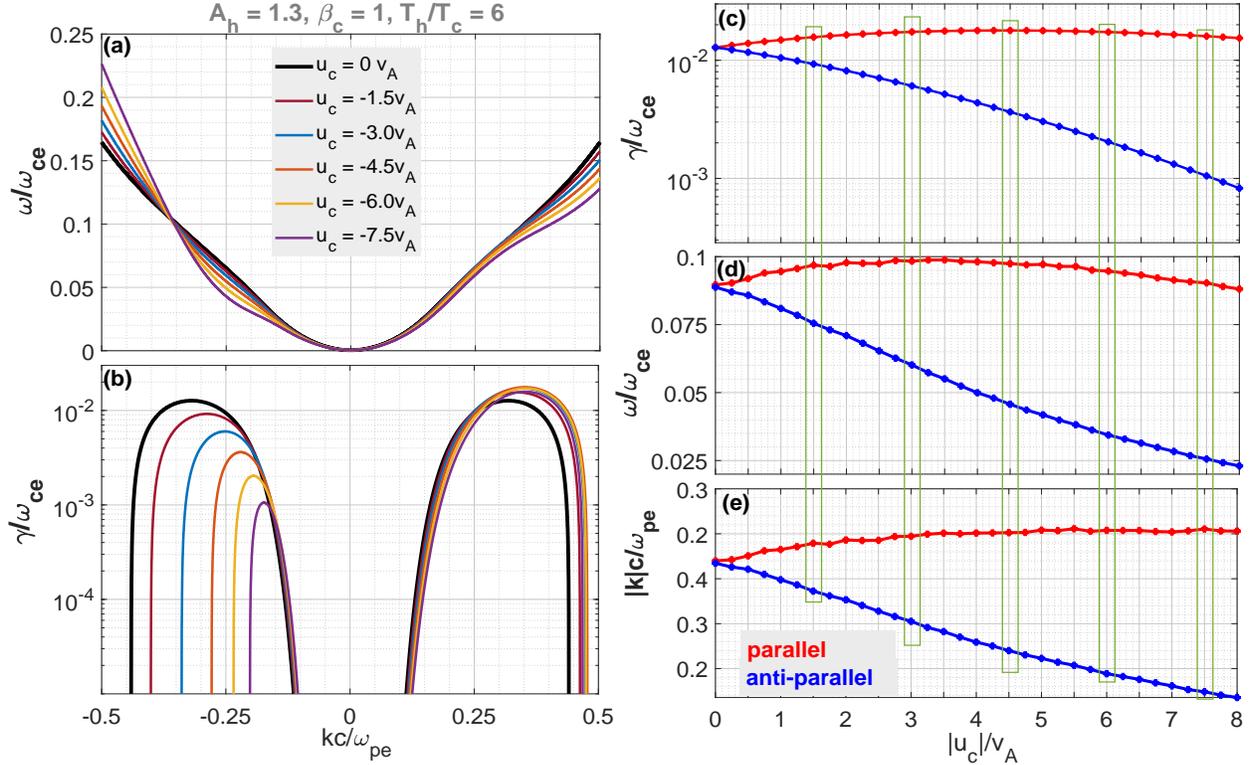}
    \caption{The results of linear stability analysis of a combined whistler heat flux and temperature anisotropy instability at fixed core electron beta and halo temperature anisotropy ($\beta_{c}=1$ and $A_{h}=1.3$), but various values of the electron heat flux set by core drift velocity $u_{c}/v_{A}$. Panels (a) and (b) present dispersion curves ($\omega/\omega_{ce}$ vs. $kc/\omega_{pe}$) and growth rates ($\gamma/\omega_{ce}$ vs. $kc/\omega_{pe}$) of whistler waves propagating parallel ($k>0$) and anti-parallel ($k<0$) to the electron heat flux, where $\omega_{ce}$ and $\omega_{pe}$ are respectively electron cyclotron and plasma frequencies. Panels (c)--(e) present the growth rate, frequency and wave number of the fastest-growing parallel and anti-parallel whistler waves at various values of core drift velocity $u_{c}/v_{A}$. The green bars in panels (c)--(e) indicate $u_{c}/v_{A}$ values used in simulation runs presented in Section \ref{sec:res_part}.}
    \label{fig1}
\end{figure*}

\begin{figure}[ht!]
    \centering
    \includegraphics[width=0.5\textwidth, height = 0.6\textheight]{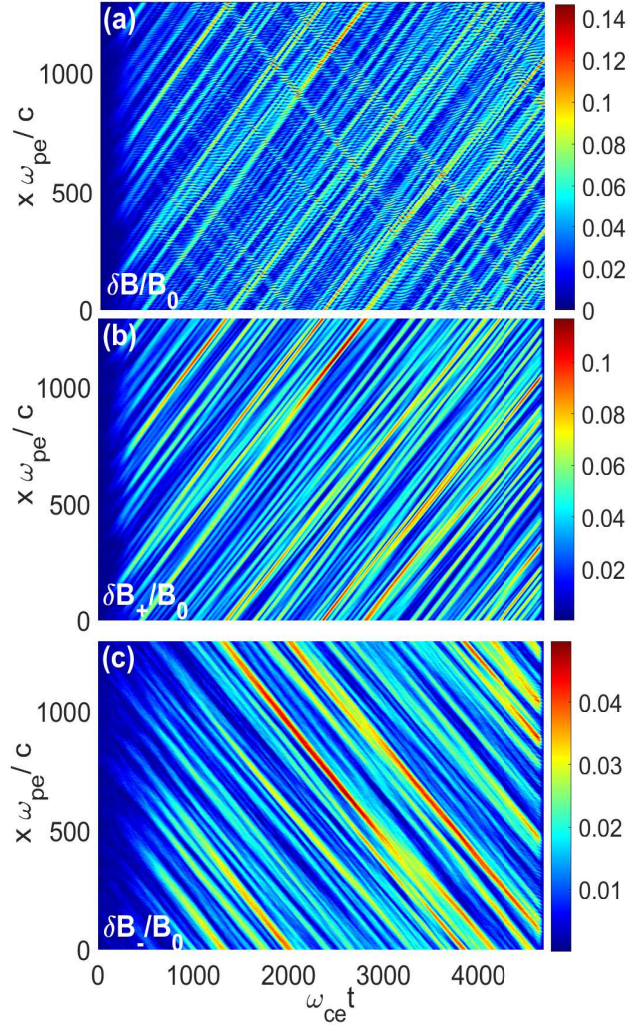}
    \caption{The results of a simulation run performed at $\beta_{c}=1$, $A_{h}=1.3$ and $u_{c}/v_{A}=-3$. Panel (a) presents the magnitude of magnetic field $\delta {\bf B}(x,t)= \delta B_y(x,t)\hat{y}+\delta B_z(x,t)\hat{z}$ perpendicular to background magnetic field $B_0\hat{x}$. Panels (b) and (c) demonstrate the magnitude of magnetic fields $\delta {\bf B}_{+}(x,t)$ and $\delta {\bf B}_{-}(x,t)$ corresponding to whistler waves propagating parallel and anti-parallel to the electron heat flux. The magnetic field fluctuations were decomposed into those propagating parallel and anti-parallel to the electron heat flux using the Fourier transform (Section \ref{sec:setup}).}
    \label{fig2}
\end{figure}

Figure \ref{fig2} presents results of a simulation run performed at $\beta_{c}=1$, $A_{h}=1.3$ and $u_{c}/v_{A}=-3$. We consider the dynamics of magnetic field $\delta {\bf B}(x,t)=\delta{B_{y}(x,t)}\hat{y}+\delta{B_{z}(x,t)}\hat{z}$ perpendicular to background magnetic field $B_0 \hat{x}$. Panel (a) presents the magnetic field magnitude $\delta B(x,t)/B_0$ and demonstrates the growth of magnetic field fluctuations propagating both parallel and anti-parallel to the electron heat flux. Using Fourier transform, $\delta{\bf B}(x,t)=\int \delta{\bf B}_{k\omega}\;e^{i(kx-\omega t)}\;dkd\omega$, we decompose magnetic field fluctuations into those propagating parallel and anti-parallel to the electron heat flux, $\delta{\bf B}(x,t)=\delta{\bf B}_{+}(x,t)+\delta{\bf B}_{-}(x,t)$, where $\delta{\bf B}_{+}(x,t)=\int_{\omega/k>0} \delta{\bf B}_{k\omega}\;e^{i(kx-\omega t)}\;dkd\omega$ and $\delta{\bf B}_{-}(x,t)=\int_{\omega/k<0} 
\delta{\bf B}_{k\omega}\;e^{i(kx-\omega t)}\;dkd\omega$. Both $\delta{\bf B}_{+}$ and $\delta{\bf B}_{-}$ have right-hand polarization (not shown here) and correspond to parallel and anti-parallel whistler waves expected based on linear stability analysis (Figure \ref{fig1}). Panels (b) and (c) show that over the computation time parallel and anti-parallel whistler waves reach peak amplitudes of about $0.1 B_0$  and $0.05B_0$, respectively.

\begin{figure*}[ht!]
    \centering
    \includegraphics[width=1\textwidth]{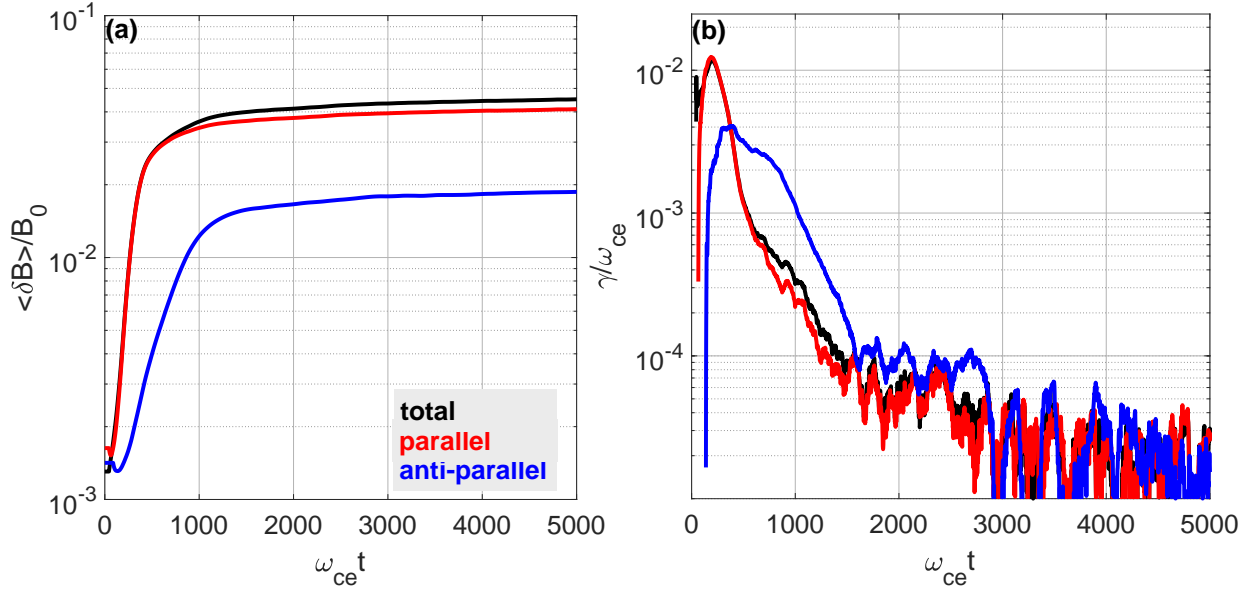}
    \caption{Panel (a) presents the temporal evolution of averaged magnetic field magnitudes of parallel and anti-parallel whistler waves observed in the simulation run shown in Figure \ref{fig2}; the magnetic field magnitudes were averaged over the simulation box, $\langle \delta B_{\pm}\rangle=\left[L^{-1}\int_0^{L}|\delta {\bf B}_{\pm}|^2\;dx\right]^{1/2}$. Panel (b) presents the corresponding growth rates, $\gamma(t)=d/dt\left[\langle\;\delta B_{\pm} \rangle\;\right]$.}
    \label{fig3}
\end{figure*}

Figure \ref{fig3} presents averaged amplitudes and growth rates of the parallel and anti-parallel whistler waves. Panel (a) shows the temporal evolution of magnetic field amplitudes averaged over the simulation box, $\langle \delta B_{\pm}\rangle=\left[L^{-1}\int_0^{L}|\delta{\bf B}_{\pm}|^2\;dx\right]^{1/2}$, and shows that within the computation time parallel and anti-parallel whistler waves saturate and the saturated amplitudes are $B_{w}^{+}/B_0\approx 0.04$ and $B_{w}^{-}/B_0\approx 0.02$. Panel (b) presents the temporal evolution of whistler wave growth rates computed as $d/dt\left[\;{\rm ln}\;\langle \delta B_{\pm}\rangle\;\right]$. The initial growth rates of parallel and anti-parallel whistler waves are respectively around 0.01 and $0.003\;\omega_{ce}$, both consistent  within a few tens of percent with linear stability results (Figure \ref{fig1}c).

\begin{figure*}[ht!]
    \centering
    \includegraphics[width=1\textwidth]{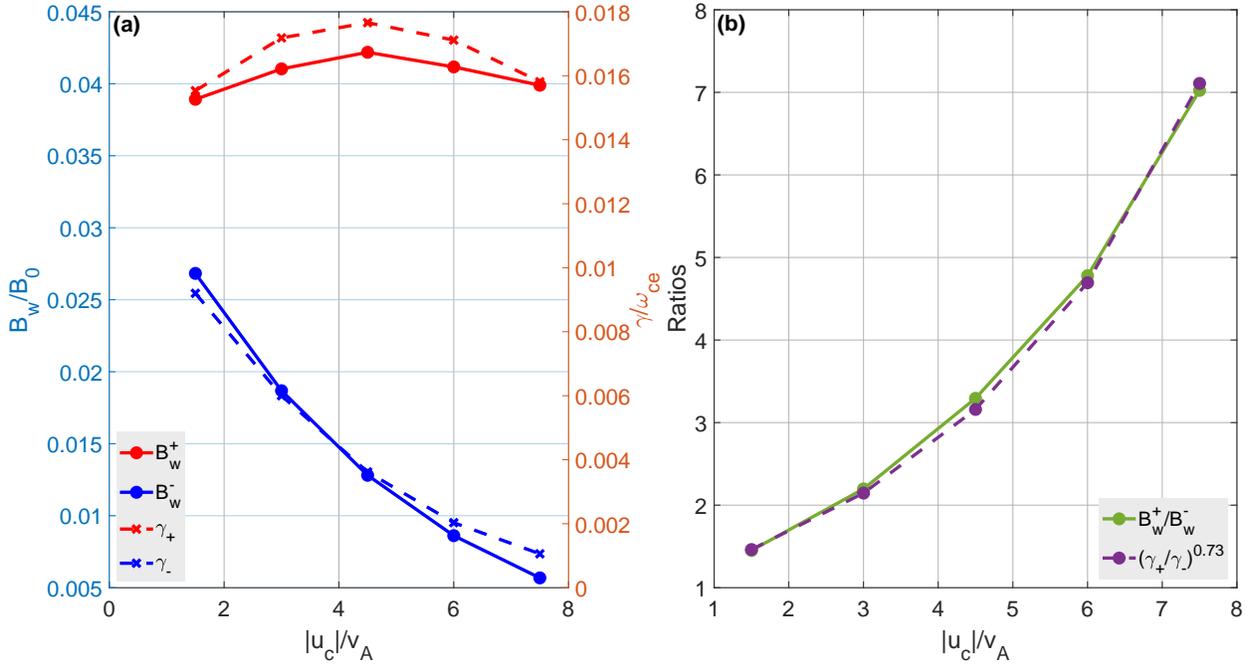}
    \caption{The results of simulation runs performed at $\beta_{c}=1$ and $A_{h}=1.3$, but various values of core electron drift velocity $u_{c}/v_{A}$ (Section \ref{sec:res_part}). Panel (a) presents saturated amplitudes $B_{w}^{+}/B_0$ and $B_{w}^{-}/B_0$ of parallel and anti-parallel whistler waves along with their {\it initial} linear growth rates $\gamma_{+}/\omega_{ce}$ and $\gamma_{-}/\omega_{ce}$ also shown in Figure \ref{fig1}. Panel (b) demonstrates that the ratios $B_{w}^+/B_{w}^-$ and $\gamma_{+}/\gamma_{-}$ are closely correlated, $B_{w}^+/B_{w}^-\approx (\gamma_+/\gamma_-)^{0.73}$.}
    \label{fig4}
\end{figure*}

Figure \ref{fig4} presents results of simulation runs performed at $\beta_{c}=1$, $A_{h}=1.3$ and various values of $u_{c}/v_{A}$ indicated in panels (c)--(e) in Figure \ref{fig1}. In all these simulation runs, parallel and anti-parallel whistler waves saturated within the computation time and we computed the averaged amplitudes $B_{w}^{+}$ and $B_{w}^{-}$ reached by the end of each simulation run. Panel (a) shows that the saturated amplitude $B_{w}^{+}$ of parallel whistler waves is around $0.04B_0$ and varies by less than several tens of percent over the considered range of $u_{c}/v_{A}$ values. In contrast, the saturated amplitude $B_{w}^{-}$ of anti-parallel whistler waves monotonously decreases from $0.025$ to $0.005B_0$. Interestingly, according to panel (a) the dependencies of the saturated amplitudes $B_{w}^{+}$ and $B_{w}^{-}$ on $u_{c}/v_{A}$ are almost identical with those of {\it initial} linear growth rates $\gamma_{+}$ and $\gamma_{-}$. Panel (b) demonstrates that the ratio $B_{w}^{+}/B_{w}^{-}$ is closely correlated with $\gamma_{+}/\gamma_{-}$ and the best power law fit is $B_{w}^{+}/B_{w}^{-}\approx (\gamma_+/\gamma_{-})^{0.73}$. This relation naturally predicts lower saturation amplitudes of anti-parallel whistler waves compared to parallel whistler waves, because the former always have lower {\it initial} linear growth rates in our model (Figure \ref{fig1}).

\begin{figure*}[ht!]
    \centering
    \includegraphics[width=1\textwidth]{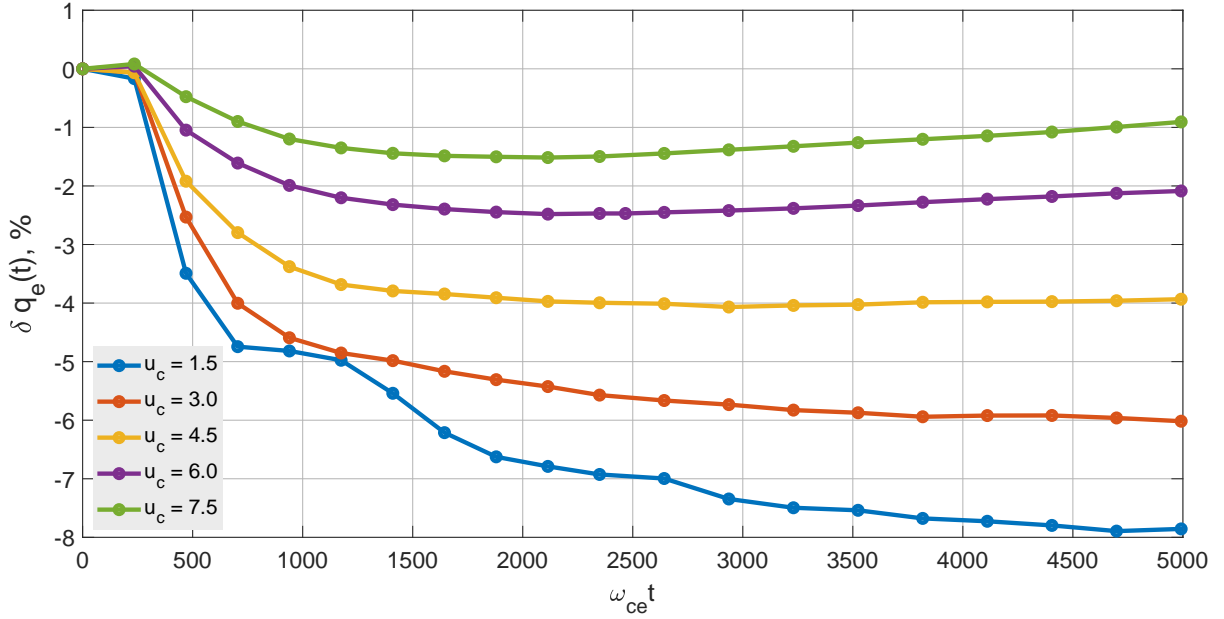}
    \caption{The temporal evolution of the electron heat flux in simulation runs performed at $\beta_{c}=1$ and $A_{h}=1.3$, but various values of core electron drift velocity $u_{c}/v_{A}$ (Section \ref{sec:res_part}). The panel presents the relative electron heat flux variation in percents, $\delta q_{e}=100\%\cdot \left[\; q_{e}(t)/q_{e}(0)-1\;\right]$, where $q_{e}(t)$ is the electron heat flux averaged over the simulation box.}
    \label{fig5}
\end{figure*}

Figure \ref{fig5} presents the temporal evolution of the electron heat flux in the considered simulation runs. We demonstrate the electron heat flux variation $\delta q_{e}(t)=\left[ q_e(t)-q_e(0)\right]/q_e(0)$ in percents, where $q_{e}(t)$ is the electron heat flux averaged over the simulation box and $q_e(0)$ is its initial value. The electron heat flux suppression is most efficient, $\delta q_{e}\approx -10\%$, in the simulation run with $u_{c}/v_{A}=-1.5$, while the efficiency drops to about 1\% at $u_{c}/v_{A}=-7.5$. There will be a natural positive correlation between the efficiency of electron heat flux suppression and the amplitude of anti-parallel whistler waves (shown further), because $B_{w}^-/B_0$ is larger for smaller values of core electron drift velocity $|u_{c}|/v_{A}$ (Figure \ref{fig4}a). Note that electron heat flux suppression of 10\% is relatively large compared to a few percent variation observed in the simulations of a pure whistler heat flux instability \citep{Kuzichev19}.

\section{Results of all simulations} \label{sec:res_all}

\begin{figure*}[ht!]
    \centering
    \includegraphics[width=1\textwidth]{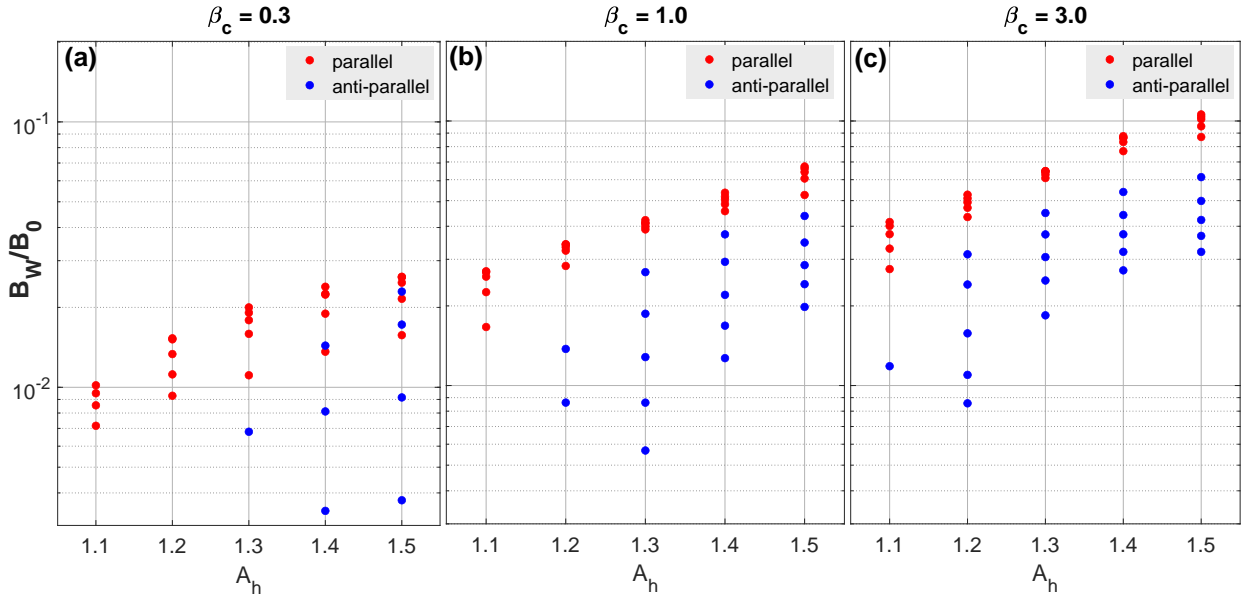}
    \caption{The results of all the 75 simulation runs performed at various values of core electron beta $\beta_{c}$, halo temperature anisotropy $A_{h}$ and core electron drift velocity $u_{c}/v_{A}$ (Table \ref{table}). Each of panels (a)--(c) presents saturated amplitudes of parallel and anti-parallel whistler waves observed in 25 simulation runs performed at $\beta_{c}=0.3, 1$ and 3. Note that we present the amplitude of a whistler wave provided that its initial linear growth rate is larger than $10^{-3}\omega_{ce}$; otherwise the computation time of $5000\omega_{ce}^{-1}$ is insufficient for the whistler waves to saturate in our simulations.}
    \label{fig6}
\end{figure*}

Figure \ref{fig6} presents averaged amplitudes $B_{w}^+$ and $B_{w}^-$ of parallel and anti-parallel whistler waves for all the 75 simulation runs (Table \ref{table}). Note that we demonstrate a whistler wave amplitude only if the initial linear growth rate of the whistler wave is larger than $10^{-3}\;\omega_{ce}$; otherwise the computation time of $5000\;\omega_{ce}^{-1}$ is insufficient for whistler waves to saturate. For this reason the number of points corresponding to parallel and anti-parallel whistler waves in panels (a)--(c) can be different and also less than 25. Panels (a)--(c) show that at a fixed core electron beta $\beta_{c}$ the saturated amplitudes are larger for larger halo temperature anisotropy $A_{h}$ and increase by a factor of a few between $A_{h}=1.1$ and 1.5. Also both parallel and anti-parallel whistler waves tend to saturate at larger amplitudes for larger core electron betas; for identical anisotropies saturated amplitudes increase by a factor of a few between $\beta_{c}=0.3$ and 3. The observed dependencies of the saturated amplitudes on the halo temperature anisotropy and other plasma parameters could be actually inferred from a more fundamental relation to be presented below.

\begin{table}[]
    \centering
    \caption{The best-fit parameters of the power-law fit given by Eq. (\ref{eq:power_laws}) between the saturated amplitudes of sunward and anti-sunward whistler waves and their initial linear growth rates. The best power-law fits are demonstrated in Figure \ref{fig7}.}
    \begin{tabular}{|c|c|c||c|c|}
    \hline
    $\beta_{c}$ & $C_{+}$ & $\nu_{+}$ & $C_{-}$ & $\nu_{-}$ \\
    \hline
    0.3  & 0.44 & 0.66 & 1.76 & 0.91\\
    \hline
    1    & 1.2 & 0.82 & 0.53 & 0.66\\
    \hline
    3    & 1.84 & 0.9 & 0.39 & 0.58\\
    \hline
    \end{tabular}
    \label{table_fit}
\end{table}

Figure \ref{fig7} shows that at every fixed core electron beta the saturated amplitudes of parallel and anti-parallel whistler waves are well-correlated with their {\it initial} linear growth rates. The observed trends can be fitted to power-law functions
\begin{eqnarray}
    B_{w}^{\pm}/B_0=C_{\pm} (\gamma_{\pm}/\omega_{ce})^{\nu_{\pm}},
    \label{eq:power_laws}
\end{eqnarray}
where the best fit parameters $C_{\pm}$ and $\nu_{\pm}$ are indicated in the panels and presented in Table \ref{table_fit}. The power-law indexes and multipliers corresponding to parallel and anti-parallel whistler waves are different and vary with core electron beta; in the considered range of core electron beta, the power law indexes vary in a relatively narrow range, $0.6\lesssim\nu_{\pm}\lesssim 0.9$.  The fundamental relation between the saturated amplitude and the initial linear growth rate naturally predicts larger amplitudes of parallel and anti-parallel whistler waves for larger anisotropies and core electron betas, because the increase of these parameters results in larger linear growth rates \citep{Vasko2020_pop}. This relation also indicates that anti-parallel whistler waves are expected to saturate at lower amplitudes than parallel whistler waves, because the presence of electron heat flux results in smaller growth rates of the former (Figure \ref{fig1}). Note that Eq. (\ref{eq:power_laws}) naturally explains the correlations reported in the previous section in Figure \ref{fig4}.


Figure \ref{fig8} demonstrates the efficiency of electron heat flux suppression quantified by the relative heat flux variation $\delta q_{e}$ reached at the saturation stage. We present $\delta q_{e}$ only if both parallel and anti-parallel whistler waves saturated within the computation time. Panel (a) shows that electron heat flux suppression is within about 10\%  at low temperature anisotropies ($A_{h}\lesssim 1.1$), but can be as large as 60\% at $A_{h}=1.5$. At a fixed halo temperature anisotropy, the electron heat flux suppression is more efficient at larger core electron betas. The efficiency of electron heat flux suppression is expected to correlate with the saturated whistler wave amplitudes, since the latter are positively correlated with both electron beta and temperature anisotropy (Figure \ref{fig5}). Panel (b) shows that electron heat flux suppression $\delta q_{e}$ is positively correlated with the saturated amplitude of anti-parallel whistler waves. The heat flux suppression is within 10\% at $B_{w}^{-}/B_0\lesssim 0.01$, but can be as large as 10--60\% at $B_{w}^-/B_0\gtrsim 0.01$. The electron heat flux suppression $\delta q_{e}$ is also correlated with $B_{w}^+/B_0$ (not shown here), but this correlation does not have direct experimental applications (see Section \ref{sec:disc}).

We address spectral properties of the whistler waves by computing power spectral densities of parallel and anti-parallel whistler waves over the saturation stage, ${\rm PSD}_{\omega}^{\pm}=\langle\left|\int \delta {\bf B}_{\pm}(x,t)e^{i\omega t}dt\right|^{2}\rangle$, where the integration is over a time period where the instability has saturated, while $\langle\cdot\rangle$ stands for spatial averaging over the simulation box. 
Gaussian fitting was done to determine the central wave frequency and the spectral width of the power spectral densities ${\rm PSD}_{\omega}^{\pm}$. Figure \ref{fig9}a demonstrates that both parallel and anti-parallel whistler waves have comparable relative spectral widths, $\Delta \omega_{\pm}/\omega_{\pm}\sim 0.3$--$0.8$. It is noteworthy that the relative spectral widths tend to be larger for larger core electron betas and positively correlated with initial linear growth rate of the whistler waves. The frequency of the saturated whistler waves is consistent within a few tens of percent with the frequency of initially fastest-growing whistler waves (not shown here).

\section{Discussion \label{sec:disc}}

We presented the first Particle-In-Cell simulations of a combined whistler heat flux and temperature anisotropy instability, which generalize our previous simulations of a pure whistler heat flux instability with isotropic halo population \citep{Kuzichev19}. In contrast to the pure whistler heat flux instability capable of producing only whistler waves propagating parallel to the electron heat flux (anti-sunward), the combined whistler heat flux and temperature anisotropy instability produces both whistler waves propagating parallel (anti-sunward) and anti-parallel (sunward) to the electron heat flux. We showed that the saturated amplitudes of the whistler waves are correlated their initial linear growth rates, $B_{w}/B_0\approx C(\gamma/\omega_{ce})^{\nu}$, where parameters $C$ and $\nu$ are a bit different for sunward and anti-sunward whistler waves. For typical solar wind conditions considered in our simulations the power-law index varies in a relatively narrow range, $0.6\lesssim\nu\lesssim 0.9$ (Table \ref{table_fit}). A similar scaling relation, though revealed using a few simulation runs, was reported for anti-sunward whistler waves produced by the pure whistler heat flux instability \citep{Kuzichev19}.

\begin{figure*}[ht!]
    \centering
    \includegraphics[width=1\textwidth]{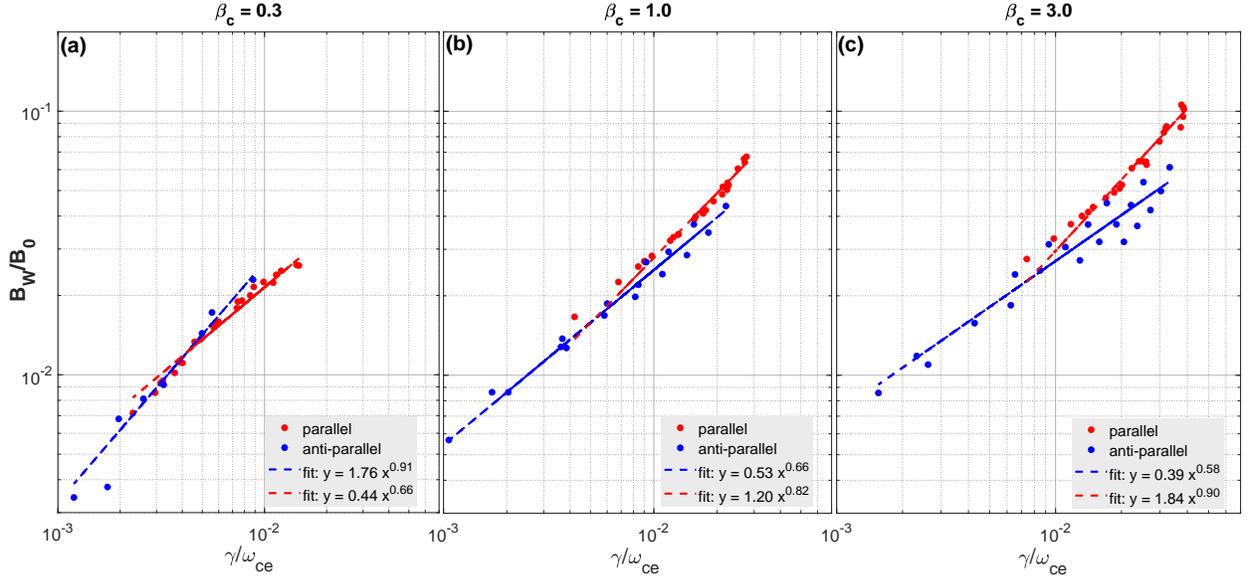}
    \caption{Each of panels (a)--(c) presents saturated amplitudes of parallel and anti-parallel whistler waves observed in 25 simulation runs performed at $\beta_{c}=0.3, 1$ and 3. The saturated amplitudes $B_{w}/B_0$ of the whistler waves are correlated with their initial linear growth rates $\gamma/\omega_{ce}$ and the best power law fits are indicated in the panels; the best fit parameters are also presented in Table \ref{table_fit}.}
    \label{fig7}
\end{figure*}

Whistler waves in our simulations saturated at $B_{w}/B_0\sim 0.01$, because we required sufficiently high initial linear growth rates, $\gamma/\omega_{ce}\gtrsim 10^{-3}$, to save computational resources (Figure \ref{fig7}). Whistler waves with such high amplitudes rarely occur in the solar wind; the observed amplitudes are typically around $10^{-3}B_0$  \cite[][]{Tong19:apj,Jagarlamudi21:aa_psp,Kretz21:aa} and, hence, correspond to initial growth rates of about $10^{-4}\omega_{ce}$. Nevertheless, we believe the revealed scaling relation is valid in a wide range of growth rates and its predictions can be compared with solar wind observations. Since linear growth rates of both sunward and anti-sunward whistler waves are larger for larger electron beta and temperature anisotropy \citep{Vasko2020_pop}, we expect whistler wave amplitudes to be positively correlated with these parameters. The corresponding positive correlations were indeed observed in the solar wind \citep{Tong19:apj}. We also showed that the whistler waves have relatively large spectral widths positively correlated with electron beta (Figure \ref{fig9}a). Spacecraft observations revealed similar values of the spectral width and its positive correlation with the electron beta \citep{Tong19:apj}. 


The revealed scaling relation (\ref{eq:power_laws}) {\it does not} correspond to whistler wave saturation via nonlinear trapping of cyclotron resonant electrons, which would occur once the bounce frequency of trapped electrons, $\omega_{T}\approx \left(kv_{\perp} eB_{w}/m_{e}c\right)^{1/2}$, is comparable to the initial linear growth rate $\gamma$, where $v_{\perp}\sim (2T_h/m_{e})^{1/2}$ is a typical perpendicular speed of resonant electrons \citep{Davidson72,Karpman74:ssr,shklyar11:angeo}. The saturation via nonlinear trapping would result in $B_{w}/B_0\propto (\gamma/\omega_{ce})^{2}$ that is inconsistent with the observed scaling relation. The nonlinear trapping does not occur when whistler waves have a sufficiently large spectral width or low amplitude such that the  resonant velocity $v_{||}=(\omega-\omega_{ce})/k$ is distributed in a range much wider than the nonlinear resonance width $\omega_{T}/k$ \citep[][]{Sag&Gal69,Karpman74:ssr}. In this case the nonlinear evolution and saturation of whistler waves could be described within the quasi-linear theory. Thus, the quasi-linear description applies provided that
\begin{eqnarray*}
\frac{\partial }{\partial \omega}\left(\frac{\omega-\omega_{ce}}{k}\right)\Delta \omega\gg \frac{\omega_{T}}{k},
\end{eqnarray*}
which can be rewritten as follows
\begin{eqnarray}
\frac{\Delta \omega}{\omega}\gg \left(\frac{B_w}{B_0}\right)^{1/2}\left(\frac{\beta\;\omega}{\omega_{ce}-\omega}\right)^{1/4},
\label{eq:qlt_crit}
\end{eqnarray}
where $\beta=\beta_{h} n_0/n_h$ and $\beta_{h}=8\pi n_h T_{h}/B_0^2$ is the halo electron beta. We computed the ratio between the left- and right-hand sides of Eq. (\ref{eq:qlt_crit}) for sunward and anti-sunward whistler waves observed in the simulations. Figure \ref{fig9}b shows that both sunward and anti-sunward whistler waves have relatively large spectral widths or low amplitudes to make quasi-linear description applicable at initial growth rates realistic of the solar wind. The same statement was shown to be valid for whistler waves actually observed in the solar wind \citep{Tong19:apj}. Note that quasi-linear description may fail at growth rates exceeding $0.01\omega_{ce}$, which are not realistic of solar wind plasma according to typically observed amplitudes of $10^{-3}B_0$. The derivation of the scaling relation (\ref{eq:power_laws}) using quasi-linear computations will be presented in a separate paper, where its validity in case of both Maxwell and $\kappa-$distributions of the halo population will be demonstrated.

The combined whistler heat flux and temperature anisotropy instability is likely operating in the solar wind. Among strong indications for its operation are statistically significant observations of the halo population with perpendicular temperature anisotropy \citep{Pierrard16,Jagarlamudi20:apj_helios,Salem21:arxiv} and preferential occurrence of whistler waves in association with isotropic or perpendicular anisotropic halo \citep{Tong19,Tong19:apj,Jagarlamudi20:apj_helios}. The operation of this instability may seem to be at conflict with recent reports of predominantly anti-sunward whistler waves in the solar wind \citep{Kretz21:aa}. In fact, the small and still uncertain occurrence of sunward whistler waves can be caused by several reasons. First, whistler waves in the solar wind are indeed produced by other instabilities including a whistler heat flux instability with isotropic or parallel-anisotropic halo electrons \citep{Tong19} and a somewhat similar instability associated with a deficit of sunward electrons \citep{Bercic21:aa}; both these instabilities produce only anti-sunward whistler waves. Second, even in the presence of a perpendicular halo temperature anisotropy, the electron heat flux breaks the symmetry between sunward and anti-sunward whistler waves resulting in smaller growth rates and, hence, smaller saturated amplitudes of the former; therefore, the presence of sunward whistler waves is more likely obscured in magnetic field spectra by solar wind turbulence.

\begin{figure*}[ht!]
    \centering
    \includegraphics[width=1\textwidth]{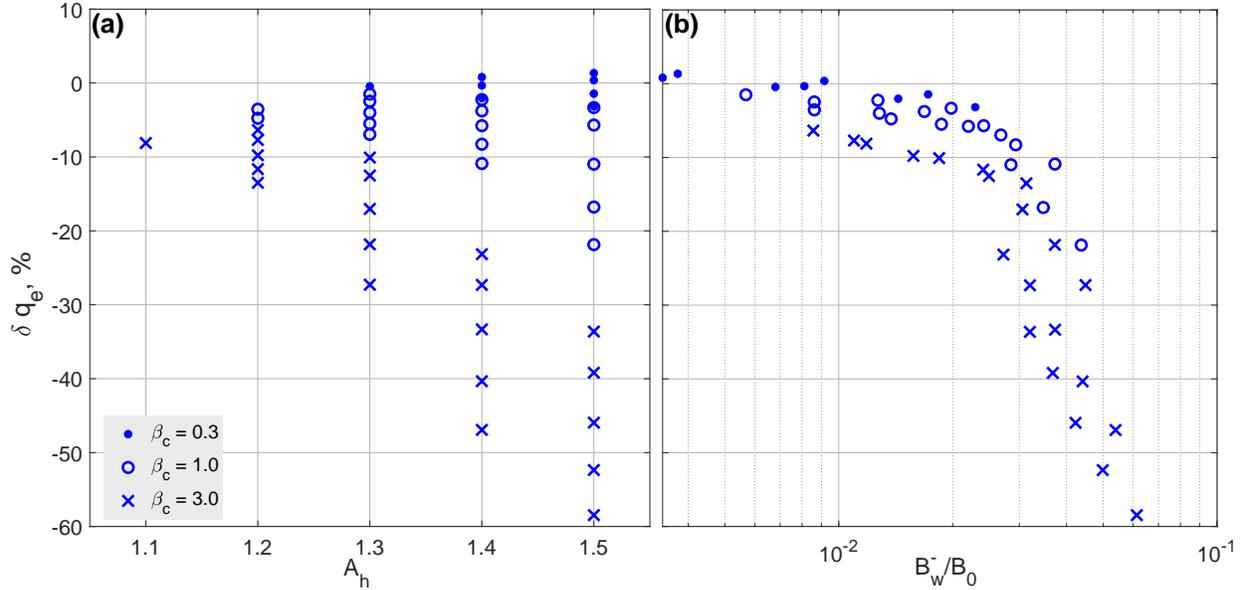}
    \caption{Panel (a) presents the relative electron heat flux variations reached by the end of our 75 simulation runs performed at various values of core electron beta $\beta_{c}$, halo temperature anisotropy $A_{h}$ and core electron drift velocity $u_{c}/v_{A}$ (Table \ref{table}). Panel (b) shows that the electron heat flux variation versus the saturated amplitude of whistler waves propagating anti-parallel to the electron heat flux (sunward whistler waves). Note that we only present results of those simulations runs, where sunward whistler waves had initial growth rates larger than $10^{-3}\omega_{ce}$ and could saturate over the computation time.}
    \label{fig8}
\end{figure*}

\begin{figure*}[ht!]
    \centering
    \includegraphics[width=1\textwidth]{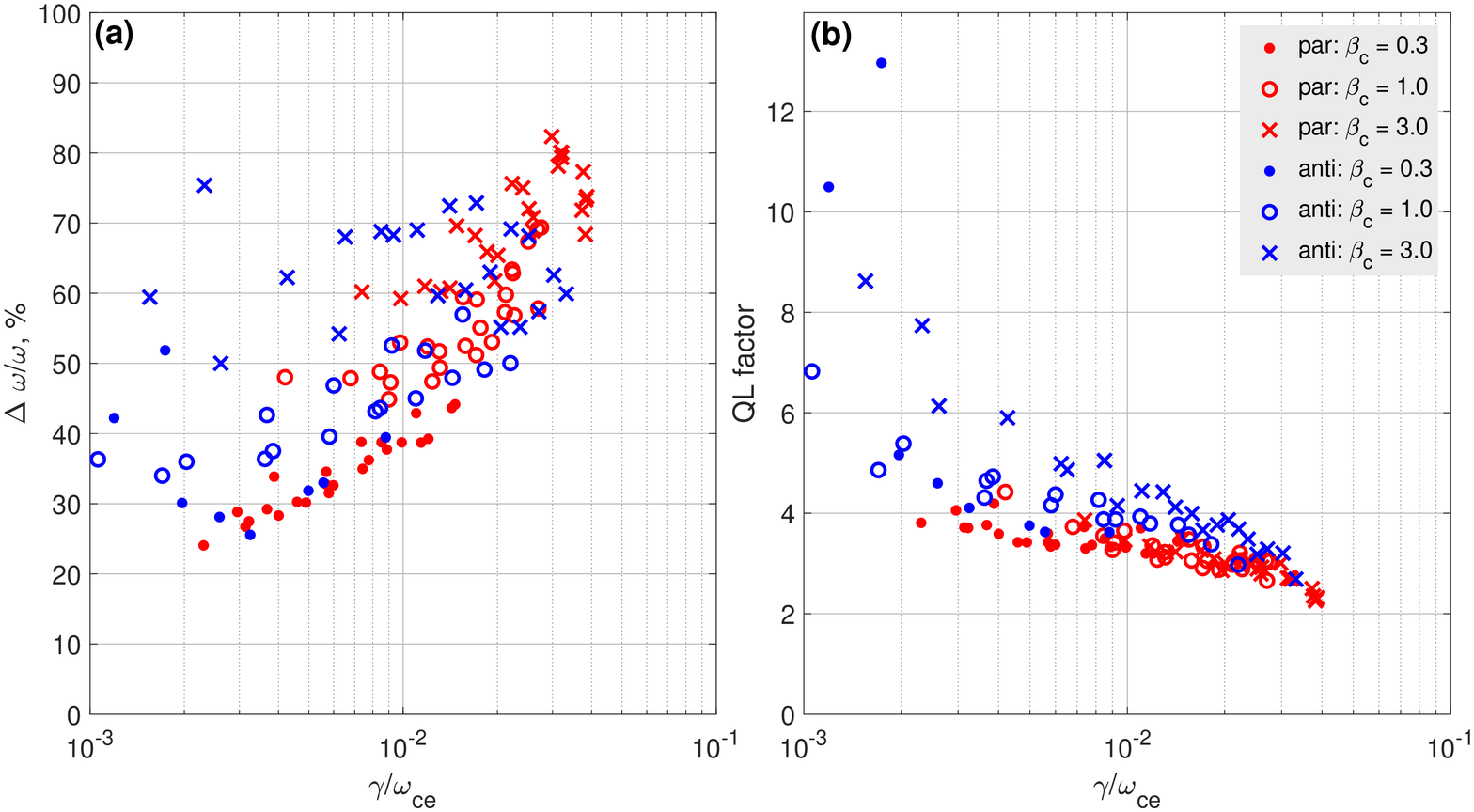}
    \caption{Panel (a) presents the relative spectral widths $\Delta \omega/\omega$ of parallel (anti-sunward) and anti-parallel (sunward)  whistler waves observed over the saturation stage in the simulations performed at various background plasma parameters (Table \ref{table}). The spectral widths $\Delta \omega$ and central frequencies $\omega$ were computed using Gaussian fittings of the spectra. Panel (b) presents the ratio between the left- and right-sides of Eq. (\ref{eq:qlt_crit}) computed separately for sunward and anti-sunward whistler waves and plotted versus their initial linear growth rates. The fact that this ratio is larger than one implies the nonlinear evolution and saturation of the whistler waves can described within quasi-linear theory.}
    \label{fig9}
\end{figure*}

Our previous simulations showed that  anti-sunward whistler waves produced by the pure whistler heat flux instability are not efficient in the electron heat flux suppression \citep{Kuzichev19}. In contrast, the combined whistler heat flux and temperature anisotropy instability can be more efficient in the electron heat flux suppression, especially at sufficiently large core electron beta and halo temperature anisotropy; at $\beta_{c}\gtrsim 3$ and $A_{h}\gtrsim 1.3$ the electron heat flux can be suppressed by up to 30--60\% (Figure \ref{fig8}a). Note that more efficient suppression of the electron heat flux at higher electron betas is consistent with spacecraft observations \citep[e.g.,][]{Gary1999b,Tong19:apj,Halekas21:aa}, while the effect of the halo anisotropy has not been addressed experimentally yet. Importantly, the efficiency of electron heat flux suppression is positively correlated with the amplitude $B_{w}/B_0$ of sunward whistler waves and the electron heat flux can be suppressed by more than 10--60\% at $B_{w}/B_0\gtrsim 0.01$ (Figure \ref{fig9}b). This correlation allows the observed amplitude of sunward whistler waves to serve as an indicator of the efficiency of electron heat flux suppression. Since whistler waves in the solar wind have typical amplitudes below a few percent of background magnetic field (a fraction of them is sunward), we expect the typical efficiency of electron heat flux suppression within about 10\%. The recent reports of sunward whistler waves with amplitudes of $0.1B_0$ \citep{Agapitov20:apj_sunward,Mozer20:jgr} indicate however that electron heat flux can be occasionally suppressed by more than 60\%. Note that the amplitude of anti-sunward whistler waves cannot similarly indicate the efficiency of electron heat flux suppression, since anti-sunward whistler waves are also produced by other instabilities proved to be inefficient in electron heat flux suppression \citep{Tong19,Kuzichev19}.

In conclusion, the combined whistler heat flux and temperature anisotropy instability is very likely operating in the solar wind and capable of producing both sunward and anti-sunward whistler waves. The presented Particle-In-Cell simulations revealed correlations between whistler wave properties (amplitude and spectral width) and various plasma parameters, which are consistent with previous solar wind observations. This instability can be efficient in electron heat flux suppression and the amplitude of sunward whistler waves can serve as an indicator of the efficiency. We expect future spacecraft measurements to reveal the occurrence and amplitude of sunward whistler waves and allow establishing the contribution of this instability to the electron heat flux suppression in the solar wind.

\section*{Acknowledgments}
The work of I.K., I.V. and A.A. was supported by NASA grants 80NSSC21K0581 and 80NSSC23K0100. We would like to acknowledge high-performance computing support from Cheyenne (doi:10.5065/D6RX99HX) provided by NCAR’s Computational and Information Systems Laboratory, sponsored by NSF grant No. 1502923. I.V. thanks the International Space Science Institute, Bern, Switzerland or supporting the working group on "Heliospheric Energy Budget: From Kinetic Scales to Global Solar Wind Dynamics".

%








\clearpage 
\newpage

\end{document}